\documentclass[sigconf,nonacm]{acmart}
\usepackage{graphicx}
\usepackage{amsmath}

\usepackage{amssymb}
\usepackage{booktabs}
\usepackage{multirow}
\usepackage{cases}
\usepackage{array}
\usepackage{glossaries-extra}
\usepackage{xspace}
\usepackage{wrapfig}
\usepackage{makecell}
\usepackage{xcolor}
\usepackage{enumitem}
\usepackage{lipsum}
\usepackage{natbib}
\usepackage{pdfpages}
\usepackage[
    type={CC},
    modifier={by-nc-nd},
    version={4.0},
    imageposition={left}
]{doclicense}

\definecolor{myred}{rgb}{0.55, 0.0, 0.0} % a dark red color
\definecolor{myblue}{rgb}{0.0, 0.0, 0.55} % a dark blue color
\definecolor{mygreen}{rgb}{0.0, 0.55, 0.0} % a dark green color

\newabbreviation{SLM}{SLM}{Spatial Light Modulator}

\newabbreviation{CGH}{CGH}{Computer-Generated Holography}
\global\long\def\CGH{\gls{CGH}\xspace}
\newabbreviation{SI}{SI}{Spatially Invariant}
\global\long\def\SI{\gls{SI}\xspace}
\newabbreviation{SV}{SV}{Spatially Varying}
\global\long\def\SV{\gls{SV}\xspace}
\newabbreviation{SA}{SA}{Spatially Adaptive}
\global\long\def\SA{\gls{SA}\xspace}
\newabbreviation{3D}{3D}{Three-Dimensional}
\global\long\def\3D{\gls{3D}\xspace}
\newabbreviation{SVF}{SVF}{ Spatially Varying Feature }
\global\long\def\SVF{\gls{SVF}\xspace}
\newabbreviation{CNN}{CNN}{Convolutional Neural Network}
\global\long\def\CNN{\gls{CNN}\xspace}
\newabbreviation{SAM}{SAM}{Spatially Adaptive Module}
\global\long\def\SAM{\gls{SAM}\xspace}

\newabbreviation{SAC}{SAC}{Spatially Adaptive Convolution}
\global\long\def\SAC{\gls{SAC}\xspace}

\newabbreviation{ASM}{ASM}{Angular Spectrum Method}
\global\long\def\ASM{\gls{ASM}\xspace}

\newabbreviation{LR}{LR}{Learning Rate}
\global\long\def\LR{\gls{LR}\xspace}

\newabbreviation{DFN}{DFN}{Dynamic Filter Networks}

\AtBeginDocument{%
  \providecommand\BibTeX{{%
    \normalfont B\kern-0.5em{\scshape i\kern-0.25em b}\kern-0.8em\TeX}}}

\makeatletter
\def\@copyrightpermission{
\doclicenseThis
}

\copyrightyear{2024}
\setcopyright{cc}

\begin{document}

%%
%% The "title" command has an optional parameter,
%% allowing the author to define a "short title" to be used in page headers.
\title{Focal Surface Holographic Light Transport using
Learned Spatially Adaptive Convolutions}

%%
%% The "author" command and its associated commands are used to define
%% the authors and their affiliations.
%% Of note is the shared affiliation of the first two authors, and the
%% "authornote" and "authornotemark" commands
%% used to denote shared contribution to the research.
\author{Chuanjun Zheng}

\affiliation{%
  \institution{University College London}
  \country{United Kingdom}}

% \email{	chuanjunzhengcs@gmail.com}

\author{Yicheng Zhan}
\affiliation{%
  \institution{University College London }
  \country{United Kingdom}\hspace{10mm}}
% \email{ucaby83@ucl.ac.uk}

\author{Liang Shi}
\affiliation{%
  \institution{Massachusetts Institute~of~Technology}
  \country{USA}}
% \email{liangs@mit.edu}

\author{Ozan Cakmakci }
\affiliation{%
  \institution{ Google}
  \country{USA}}
% \email{ozancakmakci@google.com}

\author{Kaan Ak\c{s}it}
\authornotemark[1]
\affiliation{%
  \institution{University College London}
  \country{United Kingdom}
  \authornote{denotes corresponding author}
 }

\vspace{3mm}

% \email{k.aksit@ucl.ac.uk}

% \author{Valerie B\'eranger}
% \affiliation{%
%   \institution{Inria Paris-Rocquencourt}
%   \city{Rocquencourt}
%   \country{France}
% }

% \author{Aparna Patel}
% \affiliation{%
%  \institution{Rajiv Gandhi University}
%  \streetaddress{Rono-Hills}
%  \city{Doimukh}
%  \state{Arunachal Pradesh}
%  \country{India}}

% \author{Huifen Chan}
% \affiliation{%
%   \institution{Tsinghua University}
%   \streetaddress{30 Shuangqing Rd}
%   \city{Haidian Qu}
%   \state{Beijing Shi}
%   \country{China}}

% \author{Charles Palmer}
% \affiliation{%
%   \institution{Palmer Research Laboratories}
%   \streetaddress{8600 Datapoint Drive}
%   \city{San Antonio}
%   \state{Texas}
%   \country{USA}
%   \postcode{78229}}
% \email{cpalmer@prl.com}

% \author{John Smith}
% \affiliation{%
%   \institution{The Th{\o}rv{\"a}ld Group}
%   \streetaddress{1 Th{\o}rv{\"a}ld Circle}
%   \city{Hekla}
%   \country{Iceland}}
% \email{jsmith@affiliation.org}

% \author{Julius P. Kumquat}
% \affiliation{%
%   \institution{The Kumquat Consortium}
%   \city{New York}
%   \country{USA}}
% \email{jpkumquat@consortium.net}

%%%%%%%%% ABSTRACT
\begin{abstract}
% Computer-generated holography, a set of algorithmic methods for holography, deals with identifying holograms that faithfully reconstruct Three-Dimensional scenes in holographic displays.
% %
% Traditional CGH algorithms decompose 3D scenes into multiplanes at different depth levels. 
% %
% Those methods optimize holograms using $n$ convolutions representing light transport in free space for $n$ multiple planes, which is time-consuming and limits their capability for stimulating the hologram propagated onto the non-flat focal surface. 
% %
% Although learning-based methods have been proposed for wave propagation simulation, they are still based on plane-to-plane interaction.  
% %
% In this work, we address those issues by introducing a learned focal surface beam propagation network, a small and shallow-layer network. 
% %
% At the heart of our model, we propose spatially adaptive convolution to achieve depth-varying propagation. 
% %
% Our network serves as an innovative light transport model for directly mapping complex fields onto non-flat focal surfaces and advances holography by decreasing the computational complexity for future computer-generated holography algorithms.
Computer-Generated Holography (CGH) is a set of algorithmic methods for identifying holograms that reconstruct Three-Dimensi-onal (3D) scenes in holographic displays.
CGH algorithms decompose 3D scenes into multiplanes at different depth levels and rely on simulations of light that propagated from a source plane to a targeted plane.
Thus, for $n$ planes, CGH typically optimizes holograms using $n$ plane-to-plane light transport simulations, leading to major time and computational demands.
Our work replaces multiple planes with a focal surface and introduces a learned light transport model that could propagate a light field from a source plane to the focal surface in a single inference.
Our learned light transport model leverages spatially adaptive convolution to achieve depth-varying propagation demanded by targeted focal surfaces.
The proposed model reduces the hologram optimization process up to 1.5x, which contributes to hologram dataset generation and the training of future learned CGH models.

% New sentence on inference speed benefits.
% New sentence on its potential benefit in CGH optimization.

\end{abstract}

\vspace{-5mm}

% \ccsdesc[500]{Do Not Use This Code~Generate the Correct Terms for Your Paper}
% \ccsdesc[300]{Do Not Use This Code~Generate the Correct Terms for Your Paper}
% \ccsdesc{Do Not Use This Code~Generate the Correct Terms for Your Paper}
% \ccsdesc[100]{Do Not Use This Code~Generate the Correct Terms for Your Paper}

%%
%% Keywords. The author(s) should pick words that accurately describe
%% the work being presented. Separate the keywords with commas.


%% A "teaser" image appears between the author and affiliation
%% information and the body of the document, and typically spans the
%% page.
% \begin{teaserfigure}
%   \includegraphics[width=\textwidth]{figures/systemFigure/teaserfigure.png}
%   \caption{Actual captures from our holographic display, illustrating two different focuses highlighted by the black boxes. The conventional method, as described in the literature~\cite{realisticDefocus}, optimizes the hologram using multiple planes which offer realistic defocus effects but tend to reduce sharpness in the focus regions. In contrast, our approach optimizes the hologram based on focal surfaces. This not only improves focus and defocus responses in depth-varying regions but also reduces computational complexity.
% }
%   \label{fig:fig_1}
% \end{teaserfigure}

% \received{20 February 2007}
% \received[revised]{12 March 2009}
% \received[accepted]{5 June 2009}

%%
%% This command processes the author and affiliation and title
%% information and builds the first part of the formatted document.
% \settopmatter{printacmref=false}
% \renewcommand\footnotetextcopyrightpermission[1]{}
\maketitle
% \settopmatter{printacmref=false} % Remove ACM reference format
% \renewcommand\footnotetextcopyrightpermission[1]{} % Remove copyright notice
% \pagestyle{plain} % Remove page numbers (optional, depending on your requirements)

\vspace{2mm}
\section{introduction}
% \CGH~\cite{javidi2021roadmap} is a family of algorithmic methods used to generate holographic interference patterns.
\CGH is a family of algorithmic methods used to generate holographic interference patterns.
Identifying these interference patterns using learned~\cite{shi2022end} and optimization~\cite{kavakli2023realistic}~\CGH methods require conventional simulations of light propagation from plane-to-plane~\cite{shen2006fast, matsushima2009band}.~Recently, learned proxy methods~\cite{choi2021neural,kavakli2022learned} have been proposed to replace conventional light propagation methods~\cite{shen2006fast, matsushima2009band}.
As these learned proxy methods for light propagation are trained using camera-in-the-loop strategies, they are able to capture imperfections of optical hardware, closing the gap between theoretical 
\begingroup
\setlength{\intextsep}{0.5pt}%
\setlength{\columnsep}{12pt}%
\begin{wrapfigure}{r}[0cm]{3.4cm}
\centering
\includegraphics[width=3.4cm,height=8.5cm]{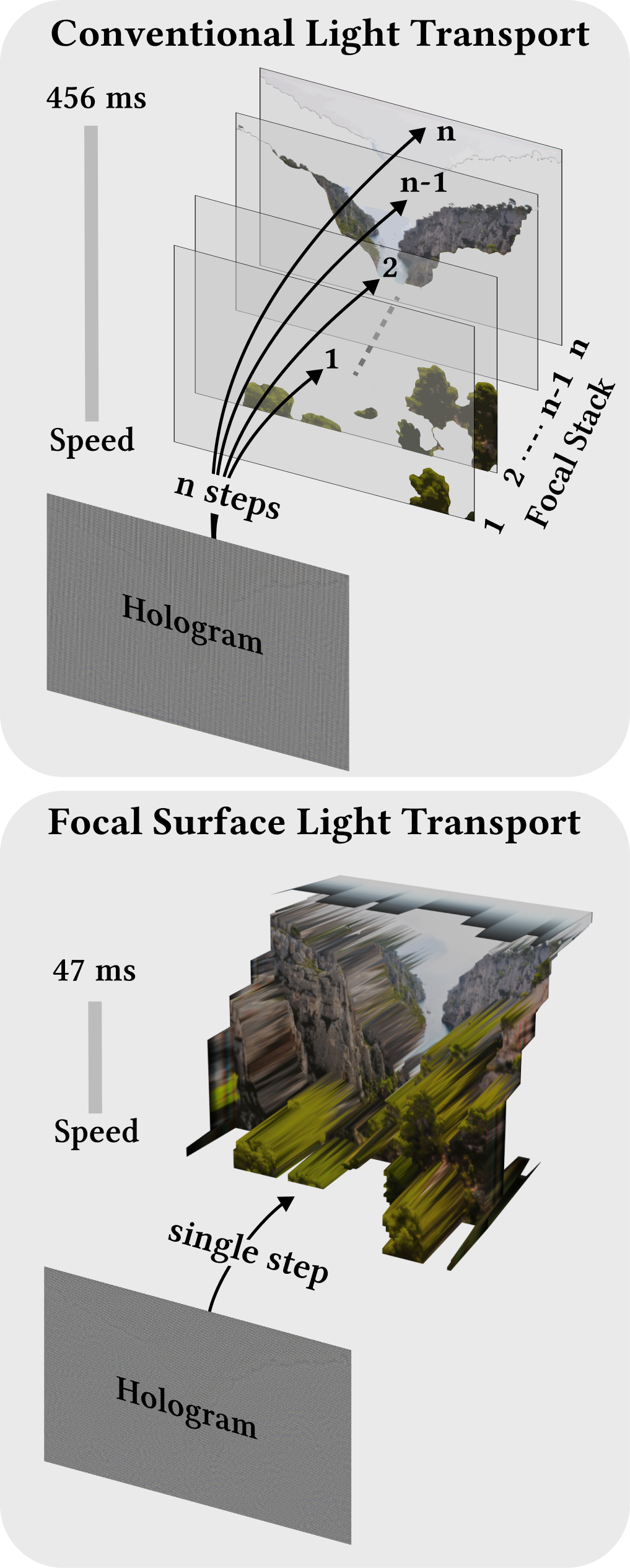}
\vspace{-7mm}
\caption{\footnotesize Conventional Light Transport VS. Proposed Focal Surface Light Transport.(Source image:~Tobi 87, Link:~\href{https://commons.wikimedia.org/wiki/File:Calanque_d'En_Vau-Cassis.jpg}{Wikimedia Commons})}
\label{fig:trad_vs_freeform}
\end{wrapfigure}
simulations and actual hardware. 
Either learned or conventional, simulating light propagation among multiple planes in a 3D volume is computationally demanding, as a 3D volume is represented with multiple planes and each plane requires a separate calculation of light propagation to reconstruct the target image.

Our work introduces a learned focal surface light propagation model that could help free light simulations from plane dependence.
Specifically, our model can propagate a phase-only hologram represented with a plane to a targeted focal surface, see Fig.~\ref{fig:trad_vs_freeform}.
In our model, we extract~\SV depth features of a focal surface by learning a set of~\SV kernels.
In addition, our model combines these~\SV learned kernels with~\SI kernels using a~\SAC.
Thus, effectively capturing~\SV and~\SI features of light propagation over a focal surface.
Our work makes the following contributions:

% \begin{itemize}[leftmargin=*]
\begin{itemize}[leftmargin=*]

\item \textbf{Learned focal surface light transport model.} By uniquely leveraging \SAC for~\CGH, we introduce a new learned light transport model. Our model identifies a mapping from a phase-only hologram represented over a plane to a targeted focal surface.
\item \textbf{Focal surface-based hologram optimization.} 
To evaluate its practicality, we utilize our model for a 3D phase-only hologram optimization application. 
Compared with conventional light propagation based hologram optimization methods~\cite{kavakli2023multi,kavakli2023realistic}, our approach accelerates the optimization process up to 1.5x, leading to speed up benefits in hologram dataset generation and training future learned~\CGH models.
\item \textbf{Experimental Validation.} We evaluate our method in simulation for various propagation distances and validate the result using a bench-top on-axis holographic display prototype.

\end{itemize}

\begin{figure*}[ht]
\captionsetup{type=figure}
\centering
\resizebox{\linewidth}{!}{
\includegraphics[width=1.0\textwidth,height=6cm]{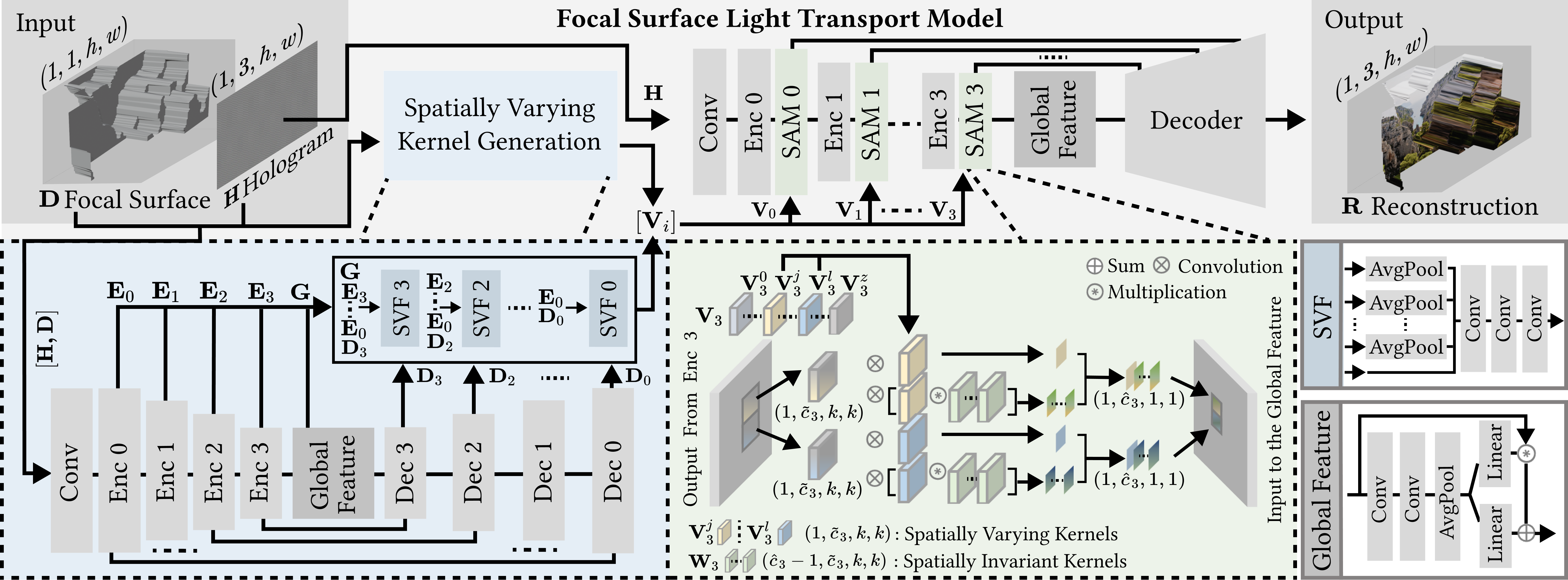}}
\vspace{-6.5mm}
\caption {Our proposed learned focal surface light transport model. The process starts with an input hologram $\mathbf{H}$ and a focal surface  $\mathbf{D}$ to generate spatially varying kernels $[\mathbf{V_{i}}]$, where $i = 0, 1, 2, 3$ indicates the index of scales. Those kernels are utilized in the Spatially Adaptive Module (SAM) to achieve focal surface light transport. In the SAM, $\mathbf{V}_{3}^{0}, \mathbf{V}_{3}^{j}, \mathbf{V}_{3}^{l}, \mathbf{V}_{3}^{z}$ represent kernels used at different spatial locations, where $0$, $j$, $l$, and $z$ indicate specific positions. (Source image:~Tobi 87, Link:~\href{https://commons.wikimedia.org/wiki/File:Calanque_d'En_Vau-Cassis.jpg}{Wikimedia Commons})}
\label{fig:system_figure}
\vspace{-2.5mm}
\end{figure*}

% \begin{figure*}[ht]
% \captionsetup{type=figure}
% \centering
% \resizebox{\linewidth}{!}{
% \includegraphics[width=1.0\textwidth]{figures/systemFigure/kernel_deatil.png}}
% \caption {}
% \label{fig:system_figure}
% \vspace{-2.5mm}
% \end{figure*}

% \input{latex/sec2-relatedworks.tex}
\section{Focal Surface Light Transport}
We introduce the~\SAC, a modified convolution structure for encoding~\SV features. Leveraging the~\SAC, our work enables the learned focal surface light transport network.

\vspace{-2mm}

\subsection{Spatially Adaptive Convolution}
\vspace{-1mm}
\subsubsection*{Standard Convolution}
Given an input feature 
$\mathbf{\tilde{I}}\in \mathbb{R}^{\tilde{c} \times \tilde{h} \times \tilde{w} }$ in a~\CNN, 
where $\tilde{c}$, $\tilde{h}$, and $\tilde{w}$ represent the number of channels, height, and width of the input $\mathbf{\tilde{I}}$ 
(in our case, $\tilde{c}=3$, $\tilde{w}=1080$, $\tilde{h}=1920$), the discrete convolution based on a~\SI kernel $\mathbf{W} \in \mathbb{R}^{\hat{c} \times \tilde{c} \times k \times k}$ is defined as:
\begin{small}
\begin{equation}
\label{eq:SI_Conv}
\underbrace{\mathbf{I}[c, x, y]}_{{\text{output}}}  = \sum_{c^{\prime}, x^{\prime}, y^{\prime} } \underbrace{\mathbf{W}[c, c^{\prime}, x^{\prime}, y^{\prime}]}_{{\text{\SI Kernel}}}  \underbrace{\mathbf{\tilde{I}}[c^{\prime}, x + x^{\prime}, y + y^{\prime}]}_{{\text{input}}},  
\end{equation}
\end{small}\noindent
where $\tilde{c}$ and $\hat{c}$ indicate the number of input and output channels. 
The indices satisfy $1 \leq c^{\prime} \leq \tilde{c}$ and $1 \leq c \leq \hat{c}$. 
The pair $(x', y')$ belongs to the set $\Omega(k)$, which specifies a $k \times k$ convolutional window.  
The summation operation acts on all input channels, which implies that each input channel contributes to every output channel. 
According to Eq.~\eqref{eq:SI_Conv}, this operation is characterized by a kernel that is spatially shared and content-independent. 
Learning-based light transport models could use Eq.~\eqref{eq:SI_Conv} as a basic operation.
However, it is challenging for this method to project a hologram onto a focal surface.
As each pixel on the hologram plane may correspond to a different depth on the focal surface, which makes the~\SI kernel a sub-optimal choice to capture~\SV features~\cite{zheng2021windowing,xu2020squeezesegv3}, including focusing or out-of-focus effects due to depth variance.
A typical solution is to employ a large number of parameters for feature encoding, resulting in an increased memory footprint.
Alternatively, we could consider using~\SV convolution~\cite{zheng2021windowing,xu2020squeezesegv3}.
The~\SV kernel $\mathbf{V}\in \mathbb{R}^{\hat{c} \times h \times w \times \tilde{c}  \times k \times k}$ incorporates two new dimensions $h,w$ into~\SI kernel, where $h$ and $w$ indicate height, and width of the output feature. 
However, relying solely on~\SV kernels may increase model capacity due to extra parameters, particularly when $h$ and $w$ are large. 
These alternative designs all demand extra network capacity. 

\vspace{-1mm}
\subsubsection*{Spatially Adaptive Convolution Operation}
% We develop a~\SA convolution to solve the aforementioned problem.
To address these problems, we utilize the~\SAC based on~\cite{xu2020squeezesegv3}.
%
% The main idea is to augment the channel of the~\SV kernels by multiplying it with standard~\SI kernels, thereby reducing the required parameters. 
Our method reduces the network parameters by multiplying the~\SV kernel with the standard~\SI kernel.
% 
% we define a~\SI kernel $\mathbf{W} \in \mathbb{R}^{\hat{c} \times \tilde{c} \times k \times k}$, where $\tilde{c}$ and $\hat{c}$ indicates the number of input and output channels.  
%
Initially, the~\SV kernel  $\mathbf{V} \in \mathbb{R}^{1 \times h \times w \times \tilde{c} \times k \times k}$ is introduced,  the output channel is set to $1$ to reduce the number of parameters.
The~\SA kernel $\mathbf{A} \in \mathbb{R}^{\hat{c} \times h \times w  \times \tilde{c}  \times k \times k} $ is computed by multiplying the $\mathbf{W}$ and $\mathbf{V}$, which defined as:
\begin{small}
\begin{equation} 
\mathbf{A}[c,x,y,c^{\prime}, x^{\prime},y^{\prime}] = \mathbf{V}[1,x,y,c^{\prime}, x^{\prime},y^{\prime}] * \mathbf{W}[c, c^{\prime},x^{\prime}, y^{\prime}],
\label{eq:sa_kernel}
\end{equation}
\end{small}\noindent
where $1\leq c\leq \hat{c}$, $1 \leq c^{\prime} \leq \tilde{c}$ ,$1 \leq  x \leq h$ and $1 \leq y \leq w$.  
Eq.~\ref{eq:sa_kernel} enhances the output channel capacity in $\mathbf{V}$  while maintaining spatially variant.
Both $\mathbf{V}$ and $\mathbf{W}$ can be either pre-defined or learned, making the network content-adaptive. By using $\mathbf{A}$,  the~\SAC is defined as:
\begin{small}
\begin{equation}
\mathbf{I}[c, x, y] = \sum_{c^{\prime},x^{\prime}, y^{\prime}} \underbrace{\mathbf{A}[c,x,y,c^{\prime}, x^{\prime},y^{\prime}]}_{{\text{SA Kernel}}} \mathbf{\tilde{I}}[c^{\prime},x + x^{\prime}, y + y^{\prime}].
\label{eq:sa_conv}
\end{equation}
\end{small}\noindent
\SAC retains both the dimensional coherence of the~\SI kernel in~\CNN and is spatially variant at the same time. 
Note that when $\mathbf{W}$ becomes an all-one tensor, Eq.~\ref{eq:sa_conv} is equivalent to the~\SV convolution in~\CNN.

\begin{figure*}[h!]
    \begin{center}

      \vspace*{-2mm}

   \begin{minipage}[t]{0.01 \linewidth} 
    \vspace*{-0.5mm}
    \begin{minipage}[t]{1 \linewidth}
     \rotatebox{90}{\footnotesize{ \biolinum Ground Truth }}
    \end{minipage} 

\vspace*{3mm}

    \begin{minipage}[t]{1 \linewidth}
     \rotatebox{90}{\footnotesize {\biolinum Ours  }}
    \end{minipage}

\vspace*{7mm}

    \begin{minipage}[t]{1 \linewidth}
     \rotatebox{90}{\footnotesize {\biolinum U-Net  }}
    \end{minipage}

    \end{minipage}
    \begin{minipage}[t]{0.97\linewidth}

    \vspace*{-4.8mm} 

        \vspace*{4.8mm} 
             \hspace{13mm}
    \begin{minipage}[t]{0.06 \linewidth}
       \vspace*{-2.3mm}
       \centerline{\footnotesize{ \biolinum  0 mm }}
   \end{minipage}
   \hspace{13mm}
   \begin{minipage}[t]{0.06 \linewidth}
       \vspace*{-2.3mm}
       \centerline{\footnotesize{ In focus }}
   \end{minipage}
   \begin{minipage}[t]{0.06 \linewidth}
       \vspace*{-2.3mm}
       \centerline{\footnotesize{   \hspace{12mm}Out of focus} }
   \end{minipage}
           \begin{minipage}[t]{0.06 \linewidth}
       \vspace*{-2.3mm}
       
       \centerline{\footnotesize{ \hspace{22mm} In focus} }
   \end{minipage}
    \hspace{28mm}
     \begin{minipage}[t]{0.06 \linewidth}
       \vspace*{-2.3mm}
       \centerline{\footnotesize{ \biolinum  10 mm }}
   \end{minipage}
    \hspace{12mm}
       \begin{minipage}[t]{0.06 \linewidth}
       \vspace*{-2.3mm}
       \centerline{\footnotesize{ In focus }}
   \end{minipage}
   \begin{minipage}[t]{0.06 \linewidth}
       \vspace*{-2.3mm}
       \centerline{\footnotesize{   \hspace{12mm}Out of focus} }
   \end{minipage}
           \begin{minipage}[t]{0.06 \linewidth}
       \vspace*{-2.3mm}
       
       \centerline{\footnotesize{ \hspace{22mm} In focus} }
   \end{minipage}
    
    \begin{minipage}[t]{1\linewidth}
       {\includegraphics[width=1\linewidth,height=4.1cm]{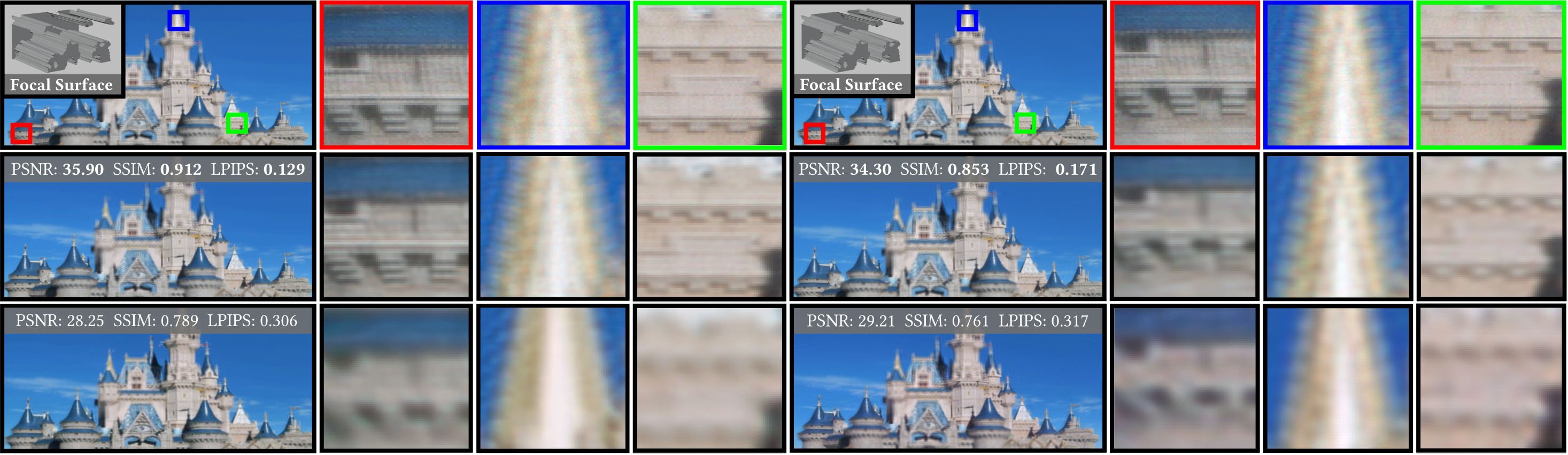}}
    \end{minipage}

    \end{minipage}

      \vspace*{-3mm}

      \caption{ 
Visual comparison of simulating light transported onto a focal surface (specified in the first row of each case) at 0 mm and 10 mm propagation distances. The ground truth is obtained via ASM~\cite{matsushima2009band}. Both focused and defocused regions indicate poor performance of the U-Net model. (Source image:~Matt H. Wade, Link:~\href{https://commons.wikimedia.org/wiki/File:Cinderella_Castle_2013_Wade.jpg}{Wikimedia Commons})}

                 \label{fig:simulation_1}
    \end{center}
 \vspace*{-4mm}
\end{figure*}

\vspace{-2mm}
\subsection{Learned Focal Surface Light Transport}
We first generate~\SV kernels to encode depth-varying features of the focal surface, which are later used in~\SAC for focal surface light transport. For the schematic figure of our system, please see Fig.~\ref{fig:system_figure}.

\vspace{-1mm}
\subsubsection*{Spatially Varying Kernel Generation}
As shown in Fig.~\ref{fig:system_figure}, the~\SV kernel generation module takes the hologram $\mathbf{H} \in \mathbb{R}^{1 \times 3 \times h \times w}$ and focal surface $\mathbf{D}\in \mathbb{R}^{1 \times 1 \times h \times w}$ as inputs. 
We adopted the architecture in RSGUNet~\cite{huang2018range} for~\SV kernel generation module. 
The output of each decoder layer is integrated with feature maps from different layers in the encoders.
Then combined features will be fed into~\SVF module to learn a set of~\SV kernels $ [ \mathbf{V}_{i} ]$, 
where 
$\mathbf{V}_{i} \in \mathbb{R}^{ n  \times \tilde{c}_{i}   \times k \times k}$, 
$i=0,1,2,3$ refers to different scale levels,
$\tilde{c}_{i}$ denotes the input channel,
$k$ is the kernel size,
and $n=\frac{h}{2^{i}} \times \frac{w}{2^{i}}$ is the number of kernels. 
The~\SVF module contains convolution layers and average pooling layers.
To mitigate artifacts, we modify the global feature module in~\cite{huang2018range} to an attention block and apply it at the bottleneck of the U-Net.

% \TODO{We modify the global feature model as an attention block that calculates global vectors as weights and biases to mitigate artifacts.}

\subsubsection*{Focal Surface Light Transport}
We leverage the generated \SV kernels to build our light transport module based on RSGUNet~\cite{huang2018range}.  
The module takes the hologram $\mathbf{H}$ as input without requiring depth, as the depth feature of the focal surface is inherently encoded within the learned~\SV kernels.
To integrate the~\SV features into the encoder, we propose a~\SAM based on~\SAC. As shown in Fig~\ref{fig:system_figure}, we first replace the~\SI kernel $\mathbf{W}$ to an all-ones tensor in Eq.~\eqref{eq:sa_kernel}, which ignores the ~\SI kernels and only considers the~\SV kernels to capture the original~\SV information.
%
% We initially replace $\mathbf{W}$ as an all-ones tensor $\mathbf{1}$ in Eq.~\eqref{eq:sa_kernel}, which means convolving the feature only with~\SV kernels to capture the original~\SV information.
%
In parallel, we introduce the~\SI kernels back to Eq.~\eqref{eq:sa_kernel} as a learning parameter and multiply with the~\SV kernels for better diverse feature extraction.
% Subsequently, we introduce~\SI kernels $\mathbf{W}$ as learning parameters in~Eq.~\eqref{eq:sa_kernel} to enhance~\SV kernels for diverse feature extraction. The output of~\SAM is the combination of those two results.
%
These features from the two operations will be concatenated to form the output of~\SAM. 
Finally, the global feature module and the decoder will process the output to generate the reconstruction at the given focal surface denoted as $\mathbf{R}$.

\vspace{-1mm}
\subsubsection*{Loss function}
We employ the $L_{2}$ norm to quantify the discrepancy between the reconstruction $\mathbf{R}$ and the target image  $\mathbf{R}^{\prime}$. Both  $\mathbf{R}$ and $\mathbf{R}^{\prime}$ are focal surface depended image reconstructions. 
Since $\mathbf{R}$ contains both focus and defocus regions~\cite{kavakli2023realistic}, we utilize a binary mask $\mathbf{M}$ that highlights only the focus parts of the image. 
The loss function for the reconstruction on a single focal surface $\mathcal{L}_{D}$ is defined as:
\begin{small}
\begin{equation} 
\mathcal{L}_{D} = \alpha_{0}\mathbf{M} \| \mathbf{R} - \mathbf{R}^{\prime} \|_{2}^{2} + \alpha_{1} (\mathbf{1}-\mathbf{M})\| \mathbf{R} - \mathbf{R}^{\prime} \|_{2}^{2},
\end{equation}
\end{small}\noindent
where $\alpha_{0}$ and $\alpha_{1}$ represent weights ($\alpha_{0}=1$ and $\alpha_{1}=0.5$).

\vspace{-2mm}
\subsection{Optimizing Holograms with Focal Surfaces} Recently, learning-based methods have been proposed to solve 3D hologram generation tasks~\cite{choi2021neural,shi2022end}. 
However, the ideal 3D hologram for the holographic display has not yet been precisely defined~\cite{kim2024holographic}. 
%
% Optimization-based solutions (need citation) could potentially help identify the optimal 3D hologram and provide target images for learned methods.
Optimization-based hologram generation methods~\cite{kavakli2023multi,kavakli2023realistic} could potentially help identify the ideal 3D hologram and generate hologram datasets for learning-based approaches.
Typically, optimization methods are based on the multiplane representation, where a full-color hologram is synthesized by making use of the phase patterns of the three color primaries.
Following previous work~\cite{kavakli2023multi}, each single-color phase pattern is obtained by:
\begin{small}
\begin{equation}
\mathbf{\hat{H}}_p \leftarrow \arg\min_{\mathbf{H_p}} \sum_{p=1}^{3} \mathcal{L} \left( \left| e^{i\mathbf{H}_p}  \otimes \mathbf{K}_p \right|^2, s\mathbf{R}_p \right), \tag{1}
\end{equation} 
\end{small}\noindent
where 
$p$ denotes the index of a color primary, 
$\mathbf{H}_p$ is the SLM phase, 
$\mathbf{\hat{H}}_p$ is the optimized SLM phase, 
$\mathbf{K}_p$ is the wavelength-dependent light transport kernel~\cite{matsushima2009band}, 
$\mathbf{R}_p$ is the target image intensity, 
$s$ is an intensity scaling factor~( $s = 1$ by default), 
$\otimes$ denotes convolution.
We substitute the conventional light transport model with our focal-surface-based model: 
\begin{small}
\begin{equation}
\mathbf{\hat{H}} \leftarrow \arg\min_{\mathbf{H}}  \mathcal{L}\left(F(\mathbf{\mathbf{H},\mathbf{D}}), s\mathbf{R}\right).
\end{equation} 
\end{small}\noindent
In this case, the hologram optimization problem is simplified.
Our approach simultaneously optimizes hologram in three color primaries and maintains phase-only at the same time.
% The phase serves as the input and the reconstruction is the output, eliminating the necessity for second-order or absolute value computations. 
%
% This approach is anticipated to enhance the fidelity of the optimization results, which will be substantiated in section 4.4.

% location effect could be also a condition, in the future we will be figure around.

\vspace*{-1mm}
\section{Evaluation and Discussion}

\begin{figure*}[h!]
    \begin{center}

\begin{minipage}[t]{1\linewidth} 

     \begin{minipage}[t]{0.5\linewidth}
     \vspace*{-1mm}
     \centerline{\footnotesize{ \biolinum  \textcolor{myblue}{Ours 6} }}

     \vspace*{-4.6mm}
         \centerline{\footnotesize{  \rotatebox{270}{ \scalebox{1.5}[1.5]{$\left\{\vphantom{\rule{0pt}{3.03cm}}\right.$}}}}    
    \end{minipage}\begin{minipage}[t]{0.5\linewidth}
     \vspace*{-1.0mm}
     \centerline{\footnotesize{ \biolinum  \textcolor{myred}{ASM 6}  }}
      \vspace*{-4.6mm}
         \centerline{\footnotesize{  \rotatebox{270}{ \scalebox{1.5}[1.5]{$\left\{\vphantom{\rule{0pt}{3.03cm}}\right.$}}}} 
    \end{minipage}
    \vspace*{-2mm}

          \begin{minipage}[t]{0.24 \linewidth}
       \vspace*{-3.0mm}
       \centerline{\footnotesize{ Rear focus }}
   \end{minipage}
    \begin{minipage}[t]{0.24 \linewidth}
       \vspace*{-3.0mm}
       \centerline{\footnotesize{ \quad Front focus} }
    \end{minipage}
           \begin{minipage}[t]{0.24 \linewidth}
       \vspace*{-3.0mm}
       \centerline{\footnotesize{ \quad \quad Rear focus} }
   \end{minipage}\hspace{0.5mm}
    \begin{minipage}[t]{0.24 \linewidth}
       \vspace*{-3.0mm}
       \centerline{\footnotesize{ \quad \quad Front focus} }
    \end{minipage}

       \vspace*{-1mm} 
    \begin{minipage}[t]{1\linewidth}
       {\includegraphics[width=1\linewidth,height=2.9cm]{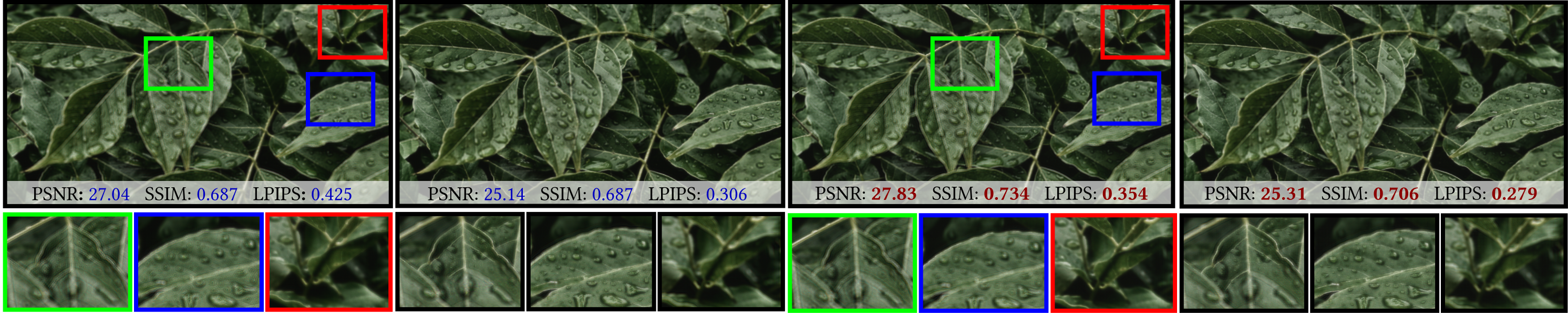}}
    \end{minipage}

       \vspace*{-1.0mm}
   \begin{minipage}[t]{0.08 \linewidth}
       \vspace*{-2.3mm}
       \centerline{\footnotesize{ Out of focus }}
   \end{minipage}
   \begin{minipage}[t]{0.08 \linewidth}
       \vspace*{-2.3mm}
       \centerline{\footnotesize{  Out of focus} }
   \end{minipage}
           \begin{minipage}[t]{0.08\linewidth}
       \vspace*{-2.3mm}
       \centerline{\footnotesize{ In focus} }
   \end{minipage}
      \begin{minipage}[t]{0.08 \linewidth}
       \vspace*{-2.3mm}
       \centerline{\footnotesize{ In focus }}
   \end{minipage}
   \begin{minipage}[t]{0.08 \linewidth}
       \vspace*{-2.3mm}
       \centerline{\footnotesize{  In focus} }
   \end{minipage}
           \begin{minipage}[t]{0.08\linewidth}
       \vspace*{-2.3mm}
       \centerline{\footnotesize{ Out of focus} }
   \end{minipage} 
      \begin{minipage}[t]{0.08 \linewidth}
       \vspace*{-2.3mm}
       \centerline{\footnotesize{ Out of focus }}
   \end{minipage}
   \begin{minipage}[t]{0.08 \linewidth}
       \vspace*{-2.3mm}
       \centerline{\footnotesize{  Out of focus} }
   \end{minipage}
           \begin{minipage}[t]{0.08\linewidth}
       \vspace*{-2.3mm}
       \centerline{\footnotesize{ In focus} }
   \end{minipage}
      \begin{minipage}[t]{0.08 \linewidth}
       \vspace*{-2.3mm}
       \centerline{\footnotesize{ In focus }}
   \end{minipage}
   \begin{minipage}[t]{0.08 \linewidth}
       \vspace*{-2.3mm}
       \centerline{\footnotesize{  In focus} }
   \end{minipage}
           \begin{minipage}[t]{0.08\linewidth}
       \vspace*{-2.3mm}
       \centerline{\footnotesize{ Out of focus} }
   \end{minipage}\hspace{0.5mm}

    \end{minipage}
         \vspace*{-3mm}
      \caption{
      Visual comparison on simulated holograms optimized using~\textcolor{myred}{ASM 6} and~\textcolor{myblue}{Ours 6} under 0 mm propagation distance. All holograms are reconstructed using ASM for evaluation. (Source image : Jaimie Phillips, Link:~\href{https://commons.wikimedia.org/wiki/File:Dewdrops_on_leaves_(Unsplash).jpg}{Wikimedia Commons})}
           
                 \label{fig:simulation_2}
    \end{center}
 \vspace*{-3mm}
\end{figure*}

We generate the focal surface light transport dataset based on previous work~\cite{kavakli2023multi,kavakli2023realistic} at the resolution $1920 \times 1080$. 
See Section 1 of the supplementary material for more details.
We use Adam optimizer  $(\beta_1=0.9, \beta_2=0.999, \alpha_{decay}= 0.5~after~50 ~epochs)$. 
The model is trained for $500$ epochs, with an initial~\LR of $2 \times 10^{-4}$.
All experiments are conducted on a single NVIDIA V100 16G GPU.

\vspace{-1mm}
\subsubsection*{Evaluation.}To assess the image quality, we utilize metrics including Peak Signal-to-noise Ratio (PSNR), Structural Similarity (SSIM), and Perceptual Similarity Metric (LPIPS)~\cite{lpips}. 
First, we assess the quality of light simulation on a focal surface. 
As shown in Tbl.~\ref{table:simulation_1}, our model outperforms U-Net~\cite{ronneberger2015u} across all metrics.
Fig.~\ref{fig:simulation_1} shows that our model preserves more high-frequency content than U-Net, providing finer details and sharper edges, closer to the ground truth.
\vspace*{3mm}
\begin{table}[h!]
\footnotesize
\setlength{\extrarowheight}{0.6pt}
\setlength{\tabcolsep}{0.6mm}
\vspace{-1mm}
\caption{ Evaluation of various light transport models on our dataset. The speed is tested by simulating an all-in-focus, full-color 3D image with six depth planes. Note that higher PSNR/SSIM and lower Params/Speed indicate better performance, denoted by $\uparrow$ and $\downarrow$ in the tables.}
\vspace*{-2mm}
\label{table:simulation_1}
\centering
\begin{tabular}{c|cccccc }
\toprule
% \multirow{1}{*}{Methods} & \multicolumn{5}{c}{Number of Image Planes } \\ \cline{2-7} 
   \multirow{2}{*}{Methods}     & PSNR (dB) $\uparrow$ & SSIM $\uparrow$   &  \multirow{2}{*}{Stage} & Params $\downarrow$      &  Speed $\downarrow$ \\

   &0 mm/10 mm &0 mm/10 mm& & (M)&(s)\\ \hline
\multirow{2}{*}{\parbox[c]{2.3cm}{  ASM (GT)~\cite{matsushima2009band} } }     &\multirow{2}{*}{ -} &\multirow{2}{*}{ -} & \multirow{2}{*}{Two} & \multirow{2}{*}{ -} & \multirow{2}{*}{0.4559} \\ 
 &                        &                   &                   &                   &                                \\
% \makecell{Learned Light \\Transport~\cite{kavakli2022learned}}& - & -& Two   &  16.588 & 0.1965 \\ 
\multirow{2}{*}{\parbox[c]{2cm}{\centering U-Net~\cite{ronneberger2015u}}}  & \multirow{2}{*}{29.662/30.112} & \multirow{2}{*}{ 0.8015/0.7760} & \multirow{2}{*}{Single}  & \multirow{2}{*}{7.7760} & \multirow{2}{*}{0.0565} \\ 
 &                        &                   &                   &                   &                                \\
Ours                           & $\mathbf{36.016}$/$\mathbf{34.279}$ & $\mathbf{0.9128}$/$\mathbf{0.8470}$ & Single  & $\mathbf{7.4446}$ & $\mathbf{0.0471}$\\  
\bottomrule
\end{tabular}
\end{table}
\vspace*{2mm}
Second, we utilize our model for a 3D phase-only hologram optimization application under $0 mm$ propagation distance.
Optimizing holograms with six target planes using~\ASM~\cite{matsushima2009band} is denoted as~\textcolor{myred}{ASM 6}, while \textcolor{mygreen}{Ours 4} and \textcolor{myblue}{Ours 6} represent optimizing holograms using our model with four and six focal surfaces, respectively.
\begin{table}[h!]
\footnotesize
\setlength{\extrarowheight}{1pt}
\setlength{\tabcolsep}{0.6mm}
\caption{ Comparison of image quality for the scene in Fig.~\ref{fig:simulation_2} among \textcolor{myred}{ASM 6}, \textcolor{myblue}{Ours 6}, and \textcolor{mygreen}{Ours 4} across different iterations at 0 mm propagation distance. Note that higher PSNR/SSIM $\uparrow$ and lower LPIPS/Speed $\downarrow$ indicate better performance.}
\vspace*{-2mm}
\label{table:capture_simulation}
\begin{tabular}{c|ccc}
\toprule
\multirow{2}{*} {\makecell{ \textcolor{myred}{ASM 6}$/$\textcolor{myblue}{Ours 6} \\ $/$ \textcolor{mygreen}{Ours 4}} }& \multicolumn{3}{c}{Iteration}   \\ \cline{2-4} 
                                & 50                      & 100                           & 200       \\  \hline
Speed (s) $\downarrow$      & \textcolor{myred}{42.580}/\textcolor{myblue}{30.182}/\textcolor{mygreen}{$\mathbf{20.869}$}   
                            & \textcolor{myred}{84.626}/\textcolor{myblue}{61.460}/\textcolor{mygreen}{$\mathbf{39.792}$}          
                            & \textcolor{myred}{171.49}/\textcolor{myblue}{119.02}/\textcolor{mygreen}{$\mathbf{77.878}$}   \\

PSNR (dB) $\uparrow$        & \textcolor{myred}{27.377}/\textcolor{myblue}{$\mathbf{27.501}$}/\textcolor{mygreen}{26.088}    
                            & \textcolor{myred}{$\mathbf{27.795}$}/\textcolor{myblue}{27.598}/\textcolor{mygreen}{26.905}          
                            & \textcolor{myred}{$\mathbf{27.801}$}/\textcolor{myblue}{27.625}/\textcolor{mygreen}{26.928}     \\

SSIM $\uparrow$     & \textcolor{myred}{$\mathbf{0.7100}$}/\textcolor{myblue}{0.6868}/\textcolor{mygreen}{0.6142}    
                    & \textcolor{myred}{$\mathbf{0.7193}$}/\textcolor{myblue}{0.6933}/\textcolor{mygreen}{0.6753}          
                    & \textcolor{myred}{$\mathbf{0.7195}$}/\textcolor{myblue}{ 0.6890}/\textcolor{mygreen}{0.6767}      \\ 
                  
LPIPS $\downarrow$  & \textcolor{myred}{$\mathbf{0.3971}$}/\textcolor{myblue}{0.4747}/\textcolor{mygreen}{0.5431}    
                    & \textcolor{myred}{$\mathbf{0.3894}$}/\textcolor{myblue}{0.4687}/\textcolor{mygreen}{0.4707}          
                    & \textcolor{myred}{$\mathbf{0.3889}$}/\textcolor{myblue}{0.4787}/\textcolor{mygreen}{0.4689}     \\ 

\bottomrule
\end{tabular}
\vspace{-5mm}
\end{table}
All holograms are reconstructed using ASM for performance assessment. As shown in Fig.~\ref{fig:simulation_2} and Tbl.~\ref{table:capture_simulation}, ~\textcolor{myblue}{Ours 6} achieves comparable results with about $70\%$ of the optimization time compared to \textcolor{myred}{ASM 6}.
Actual captures of~\textcolor{myblue}{Ours 6} and~\textcolor{myred}{ASM 6} in Fig.~\ref{fig:capture} demonstrate the capability of our model for generating 3D holograms.
For more details on the display prototype and comparisons, see  Sections 2 and 3 in supplementary material.

\vspace*{5mm}
\begin{figure}[h!]
    \begin{center}
    \vspace*{-3mm}
    
\begin{minipage}[t]{1\linewidth} 
   
    \begin{minipage}[t]{1\linewidth}
       {\includegraphics[width=1\linewidth,height=2cm]{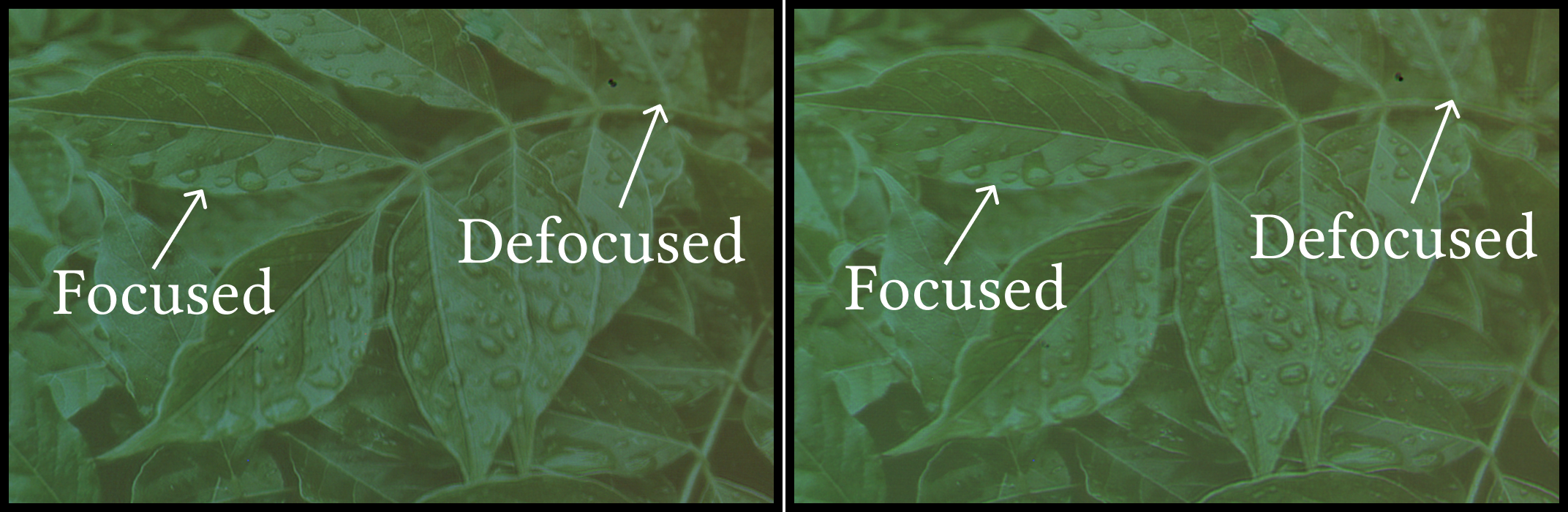}}
    \end{minipage}

       \begin{minipage}[t]{0.47\linewidth}
       \vspace*{-2.5mm}
       \centerline{\footnotesize{ \quad \quad \textcolor{myblue}{Ours 6}}}
   \end{minipage}
   \begin{minipage}[t]{0.47 \linewidth}
       \vspace*{-2.5mm}
       \centerline{\footnotesize{ \quad \quad \textcolor{myred}{ASM 6}  }  }
    \end{minipage}
    
    \end{minipage}
         \vspace*{-4mm}    
      \caption{ Comparing experimental captures of ~\textcolor{myred}{ASM 6} and~\textcolor{myblue}{Ours 6} under 0 mm propagation distances. (Source image : Jaimie Phillips, Link:~\href{https://commons.wikimedia.org/wiki/File:Dewdrops_on_leaves_(Unsplash).jpg}{Wikimedia Commons})}
           
                  \label{fig:capture}
    \end{center}
 \vspace*{-2mm}
\end{figure}

\subsubsection*{Computational Complexity Analysis.}
First, we assess the computational complexity of simulating a full-color, all-in-focus 3D image across six depth planes.
As shown in Tbl.~\ref{table:simulation_1}, conventional ASM-based model~\cite{matsushima2009band} requires eighteen forward passes to simulate a full-color, all-in-focus 3D image with six depth planes.
In contrast, our model simulates the three color-primary images simultaneously onto a focal surface with a single forward pass, reducing simulation time by 10x and achieving better image quality with fewer parameters compared to U-Net~\cite{ronneberger2015u}.
Second, we evaluate hologram optimization.
In Tbl.~\ref{table:capture_simulation}, using four focal surfaces (\textcolor{mygreen}{Ours 4}) to approximate six planes for focus and defocus guidance, speeding up optimization by up to 2x. 
Increasing the number of focal surfaces to six (\textcolor{myblue}{Ours 6}) achieves comparable results with about a 1.5x speedup.

\vspace*{-1mm}
\subsubsection*{Limitations and Future Works}
As shown in Fig.~\ref{fig:simulation_1}, the performance of our model degrades at a long propagation distance $(10 \ mm)$ compared to zero distance $(0 \ mm)$. See Section 3 in the supplementary material for more comparisons.
Future improvements could include using a factorized larger kernel for long-distance propagation. 
In addition, our model focuses on depth-varying propagation within a 3D volume, more investigation is needed for depth-varying propagation of the entire volume using conditional networks.

\begin{acks}
The authors thank reviewers for their valuable feedback; Louise L. Xie for her feedback on the manuscript; Ziyang Chen and Doğa Yılmaz for their discussions and assistance. We also thank Northeastern University for computing resources for early experiments.
\end{acks}

%%
%% The next two lines define the bibliography style to be used, and
%% the bibliography file.
\bibliographystyle{ACM-Reference-Format}
\bibliography{ref}

\AtEndDocument{\includepdf[pages=-]{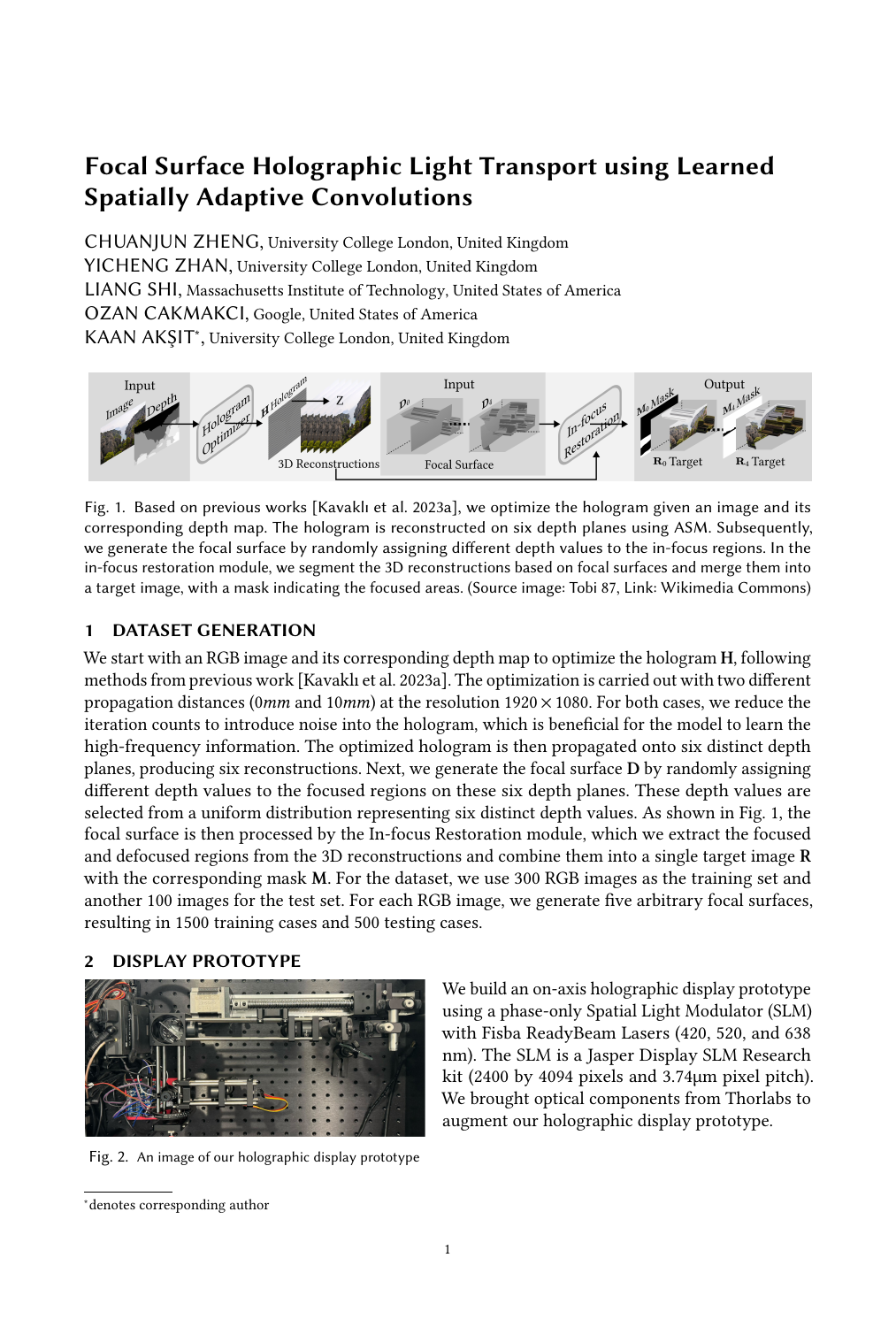}}
\end{document}

%%
%% The "title" command has an optional parameter,
%% allowing the author to define a "short title" to be used in page headers.
\title{Focal Surface Holographic Light Transport using
Learned Spatially Adaptive Convolutions}

%%
%% The "author" command and its associated commands are used to define
%% the authors and their affiliations.
%% Of note is the shared affiliation of the first two authors, and the
%% "authornote" and "authornotemark" commands
%% used to denote shared contribution to the research.

\author{Chuanjun Zheng}

\affiliation{%
  \institution{University College London}
  \country{United Kingdom}}
% \email{	chuanjunzhengcs@gmail.com}

\author{Yicheng Zhan}
\affiliation{%
  \institution{University College London}
  \country{United Kingdom}}
% \email{ucaby83@ucl.ac.uk}

\author{Liang Shi}
\affiliation{%
  \institution{Massachusetts Institute of Technology}
  \country{United States of America}}
% \email{liangs@mit.edu}

\author{Ozan Cakmakci }
\affiliation{%
  \institution{ Google}
  \country{United States of America}}
% \email{ozancakmakci@google.com}

\author{Kaan Ak\c{s}it}

\authornotemark[1]
\affiliation{%
  \institution{University College London}
  \country{United Kingdom}
  \authornote{denotes corresponding author}}

\makeatletter
\let\@authorsaddresses\@empty
\makeatother
\maketitle
\renewcommand{\shortauthors}{Zheng, C. et al.}

\vspace{-4mm}

\begin{figure*}[ht]
\captionsetup{type=figure}
\centering
\resizebox{\linewidth}{!}{
\includegraphics[width=1.0\textwidth]{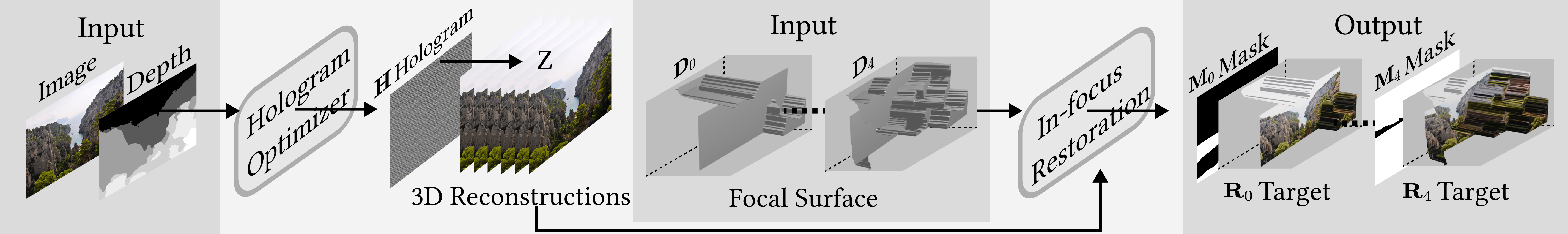}}
\vspace{-5mm}
\caption { Based on previous works~[Kavaklı et al. 2023a], we optimize the hologram given an image and its corresponding depth map. The hologram is reconstructed on six depth planes using ASM. Subsequently, we generate the focal surface by randomly assigning different depth values to the in-focus regions. In the in-focus restoration module, we segment the 3D reconstructions based on focal surfaces and merge them into a target image, with a mask indicating the focused areas. (Source image:~Tobi 87, Link:~\href{https://commons.wikimedia.org/wiki/File:Calanque_d'En_Vau-Cassis.jpg}{Wikimedia Commons})}
\label{fig:dataset}
\vspace{-2.5mm}
\end{figure*}
\vspace{-2mm}
\section{Dataset Generation}
We start with an RGB image and its corresponding depth map to optimize the hologram  $\mathbf{H}$, following methods from previous work~[Kavaklı et al. 2023a].
The optimization is carried out with two different propagation distances ($0 mm$ and $10 mm$) at the resolution $1920 \times 1080$. 
For both cases, we reduce the iteration counts to introduce noise into the hologram, which is beneficial for the model to learn the high-frequency information.
The optimized hologram is then propagated onto six distinct depth planes, producing six reconstructions. 
Next, we generate the focal surface  $\mathbf{D}$ by randomly assigning different depth values to the focused regions on these six depth planes. 
These depth values are selected from a uniform distribution representing six distinct depth values.
As shown in Fig. 1, the focal surface is then processed by the In-focus Restoration module, which we extract the focused and defocused regions from the 3D reconstructions and combine them into a single target image $\mathbf{R}$ with the corresponding mask $\mathbf{M}$.
For the dataset, we use 300 RGB images as the training set and another 100 images for the test set.
For each RGB image, we generate five arbitrary focal surfaces, resulting in 1500 training cases and 500 testing cases.

\section{Display Prototype}
\begin{wrapfigure}[7]{l}{6.5cm}
\centering
\vspace{-5mm}
\includegraphics[width=6.5cm,height=3cm]{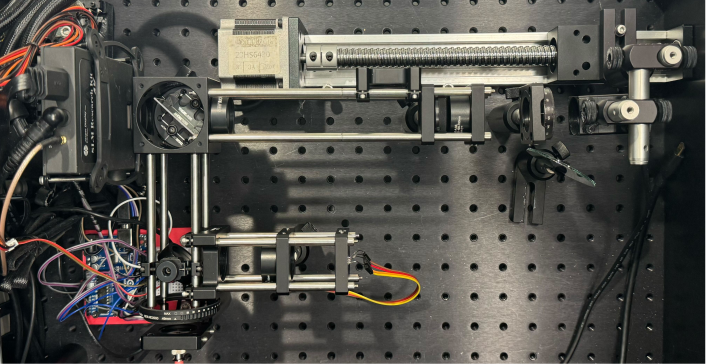}
\hspace{-2mm}
\vspace{-2mm}
\caption{\footnotesize An image of our holographic display prototype}
\label{fig:holographic_display}
\end{wrapfigure}

We build an on-axis holographic display prototype using a phase-only~\SLM with Fisba ReadyBeam Lasers (420, 520, and 638 nm).  
The~\SLM is a Jasper Display SLM Research kit (2400 by 4094 pixels and 3.74µm pixel pitch).
We brought optical components from Thorlabs to augment our holographic display prototype.

\newpage

\section{Visual Results}

\vspace{-3mm}
\begin{figure*}[h!]
    \begin{center}

   % \begin{minipage}[t]{0.19 \linewidth}
   %     \vspace*{-3mm}
   %     \centerline{\footnotesize (d) (Ours)}
   %  \end{minipage}\hspace{-0.3mm}
   %  \begin{minipage}[t]{0.19 \linewidth}
   %     \vspace*{-3mm}
   %     \centerline{\footnotesize (e) Ground Truth}
   %  \end{minipage}
   
    \begin{minipage}[t]{0.01 \linewidth} 
    \vspace*{7mm}
    \begin{minipage}[t]{1 \linewidth}
     \rotatebox{90}{\footnotesize{ \biolinum Simulations-Front Focus }}
    \end{minipage} 

\vspace*{9mm}

    \begin{minipage}[t]{1 \linewidth}
     \rotatebox{90}{\footnotesize {\biolinum Simulations-Mid Focus  }}
    \end{minipage}

\vspace*{10mm}

    \begin{minipage}[t]{1 \linewidth}
     \rotatebox{90}{\footnotesize {\biolinum Simulations-Rear Focus  }}
    \end{minipage}
    
\vspace*{5mm}

    % \begin{minipage}[t]{1 \linewidth}
    %  \rotatebox{90}{\footnotesize {\biolinum Actual Captures  }}
    % \end{minipage}

    \end{minipage}\hspace{1.3mm}\begin{minipage}[t]{0.975\linewidth} 

     \begin{minipage}[t]{0.5\linewidth}
     \centerline{\footnotesize{ \biolinum  0 mm }}

     \vspace*{-4.2mm}
         \centerline{\footnotesize{  \rotatebox{270}{ \scalebox{1.5}[1.5]{$\left\{\vphantom{\rule{0pt}{2.35cm}}\right.$}}}}    
    \end{minipage}\begin{minipage}[t]{0.5\linewidth}
     \centerline{\footnotesize{ \biolinum  10 mm }}
      \vspace*{-4.2mm}
         \centerline{\footnotesize{  \rotatebox{270}{ \scalebox{1.5}[1.5]{$\left\{\vphantom{\rule{0pt}{2.35cm}}\right.$}}}} 
    \end{minipage}
    \vspace*{-3mm}

      \hspace{0.2mm}
       \begin{minipage}[t]{0.24 \linewidth}
       \vspace*{-2.5mm}
       \centerline{\scriptsize{ \textcolor{myblue}{Ours 6} }}
   \end{minipage}
   \begin{minipage}[t]{0.24 \linewidth}
       \vspace*{-2.5mm}
       \centerline{\scriptsize{  \textcolor{myred}{ASM 6}}  }
    \end{minipage}
    \begin{minipage}[t]{0.24 \linewidth}
       \vspace*{-2.5mm}
       \centerline{\scriptsize{ \quad \textcolor{myblue}{Ours 6}} }
    \end{minipage}
           \begin{minipage}[t]{0.24 \linewidth}
       \vspace*{-2.5mm}
       \centerline{\scriptsize{ \quad \textcolor{myred}{ASM 6}} }
   \end{minipage}

    \begin{minipage}[t]{1\linewidth}
       {\includegraphics[width=1\linewidth]{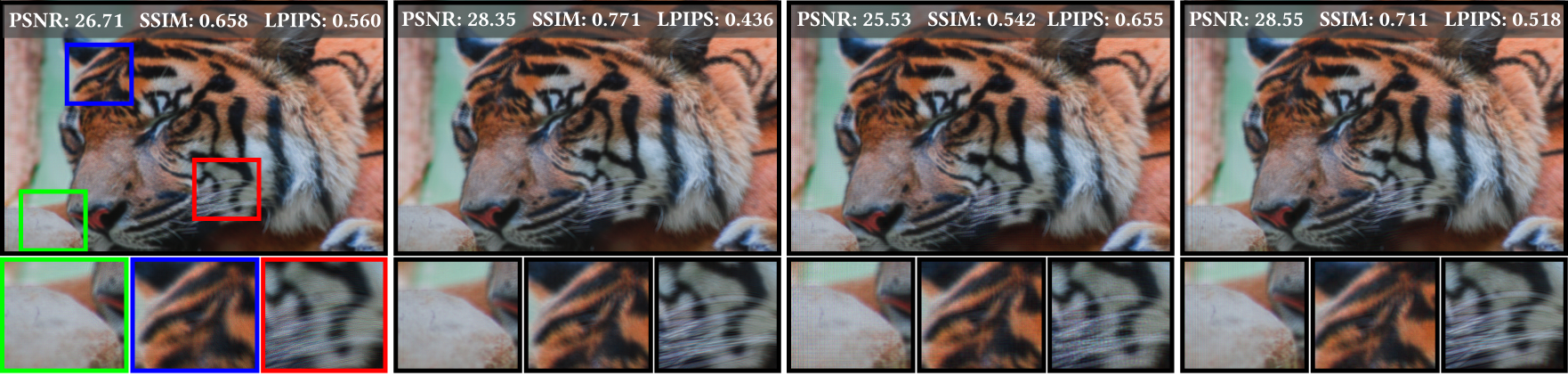}}
    \end{minipage}

 \vspace*{-1.3mm} 
   \begin{minipage}[t]{0.08\linewidth}
       \vspace*{-2.3mm}
       \centerline{\scriptsize{In focus }}
   \end{minipage}
   \begin{minipage}[t]{0.079 \linewidth}
       \vspace*{-2.3mm}
       \centerline{\scriptsize{Defocus} }
   \end{minipage}
    \begin{minipage}[t]{0.079\linewidth}
       \vspace*{-2.3mm}
       \centerline{\scriptsize{Defocus} }
   \end{minipage}
   \begin{minipage}[t]{0.079 \linewidth}
       \vspace*{-2.3mm}
       \centerline{\scriptsize{ In focus }}
   \end{minipage}
   \begin{minipage}[t]{0.079 \linewidth}
       \vspace*{-2.3mm}
       \centerline{\scriptsize{  Defocus} }
   \end{minipage}
   \begin{minipage}[t]{0.079\linewidth}
       \vspace*{-2.3mm}
       \centerline{\scriptsize{ Defocus} }
   \end{minipage}
   \begin{minipage}[t]{0.079 \linewidth}
       \vspace*{-2.3mm}
       \centerline{\scriptsize{ In focus }}
   \end{minipage}
   \begin{minipage}[t]{0.079 \linewidth}
       \vspace*{-2.3mm}
       \centerline{\scriptsize{  Defocus} }
   \end{minipage}
   \begin{minipage}[t]{0.079\linewidth}
       \vspace*{-2.3mm}
       \centerline{\scriptsize{ Defocus} }
   \end{minipage} 
   \begin{minipage}[t]{0.079 \linewidth}
       \vspace*{-2.3mm}
       \centerline{\scriptsize{ In focus }}
   \end{minipage}
   \begin{minipage}[t]{0.079 \linewidth}
       \vspace*{-2.3mm}
       \centerline{\scriptsize{ Defocus} }
   \end{minipage}
   \begin{minipage}[t]{0.079 \linewidth}
       \vspace*{-2.3mm}
       \centerline{\scriptsize{ Defocus} }
   \end{minipage}

        \vspace*{2.3mm} 
       \begin{minipage}[t]{1\linewidth}
       {\includegraphics[width=1\linewidth]{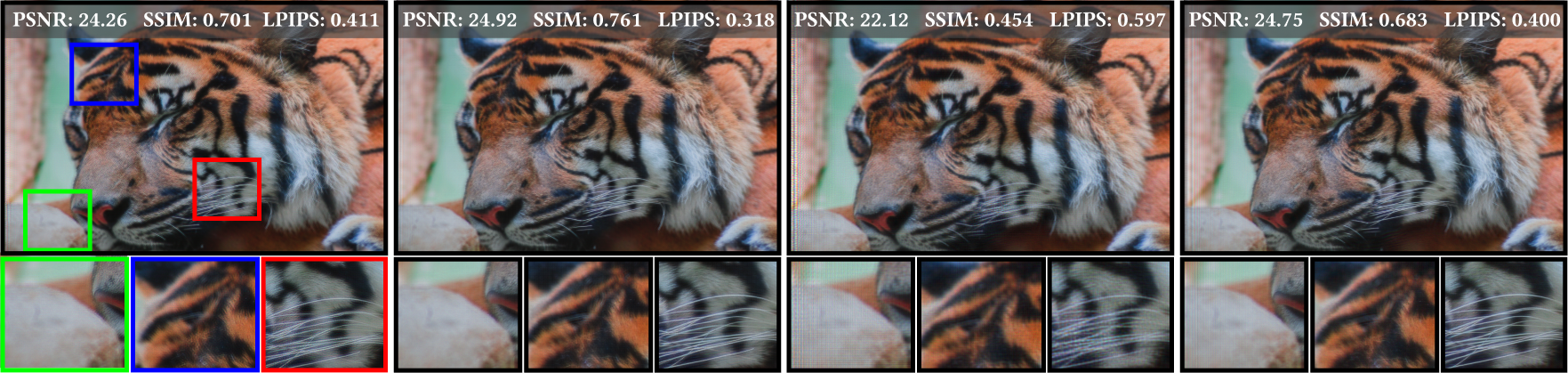}}
    \end{minipage}

   \vspace*{-1.0mm} 
   \begin{minipage}[t]{0.08\linewidth}
       \vspace*{-2.3mm}
       \centerline{\scriptsize{Defocus }}
   \end{minipage}
   \begin{minipage}[t]{0.079 \linewidth}
       \vspace*{-2.3mm}
       \centerline{\scriptsize{Defocus} }
   \end{minipage}
    \begin{minipage}[t]{0.079\linewidth}
       \vspace*{-2.3mm}
       \centerline{\scriptsize{In focus} }
   \end{minipage}
   \begin{minipage}[t]{0.079 \linewidth}
       \vspace*{-2.3mm}
       \centerline{\scriptsize{ Defocus }}
   \end{minipage}
   \begin{minipage}[t]{0.079 \linewidth}
       \vspace*{-2.3mm}
       \centerline{\scriptsize{  Defocus} }
   \end{minipage}
   \begin{minipage}[t]{0.079\linewidth}
       \vspace*{-2.3mm}
       \centerline{\scriptsize{ In focus} }
   \end{minipage}
   \begin{minipage}[t]{0.079 \linewidth}
       \vspace*{-2.3mm}
       \centerline{\scriptsize{ Defocus }}
   \end{minipage}
   \begin{minipage}[t]{0.079 \linewidth}
       \vspace*{-2.3mm}
       \centerline{\scriptsize{  Defocus} }
   \end{minipage}
   \begin{minipage}[t]{0.079\linewidth}
       \vspace*{-2.3mm}
       \centerline{\scriptsize{ In focus} }
   \end{minipage} 
   \begin{minipage}[t]{0.079 \linewidth}
       \vspace*{-2.3mm}
       \centerline{\scriptsize{ Defocus }}
   \end{minipage}
   \begin{minipage}[t]{0.079 \linewidth}
       \vspace*{-2.3mm}
       \centerline{\scriptsize{ Defocus} }
   \end{minipage}
   \begin{minipage}[t]{0.079 \linewidth}
       \vspace*{-2.3mm}
       \centerline{\scriptsize{ In focus} }
   \end{minipage}

      \vspace*{2.3mm}

        \begin{minipage}[t]{1\linewidth}
       {\includegraphics[width=1\linewidth]{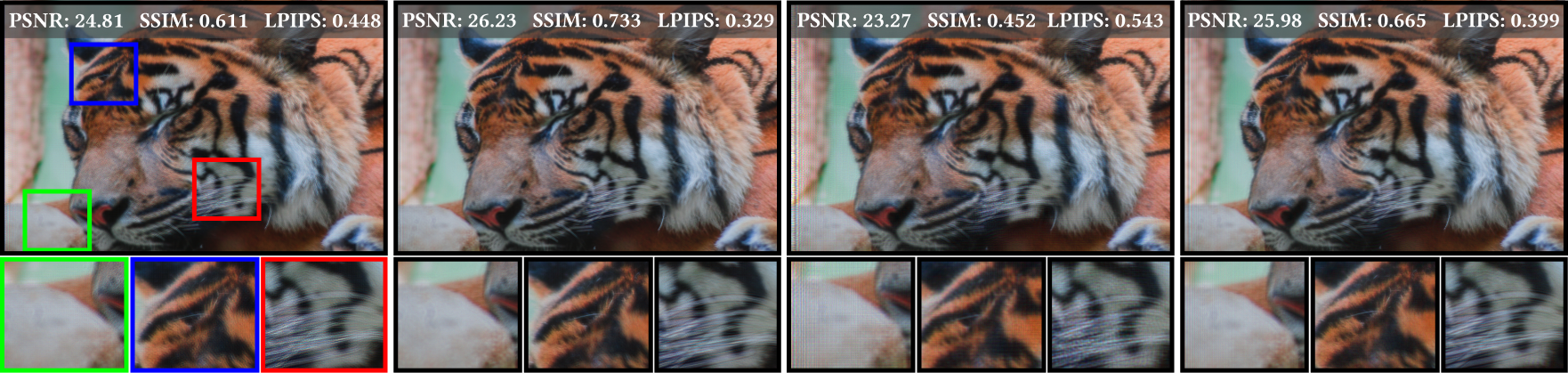}}
    \end{minipage}

   \vspace*{-1.0mm}
   \begin{minipage}[t]{0.083 \linewidth}
       \vspace*{-2.3mm}
       \centerline{\scriptsize{ Defocus }}
   \end{minipage}
   \begin{minipage}[t]{0.079 \linewidth}
       \vspace*{-2.3mm}
       \centerline{\scriptsize{In focus} }
   \end{minipage}
    \begin{minipage}[t]{0.079\linewidth}
       \vspace*{-2.3mm}
       \centerline{\scriptsize{Defocus} }
   \end{minipage}
   \begin{minipage}[t]{0.079 \linewidth}
       \vspace*{-2.3mm}
       \centerline{\scriptsize{ Defocus }}
   \end{minipage}
   \begin{minipage}[t]{0.079 \linewidth}
       \vspace*{-2.3mm}
       \centerline{\scriptsize{  In focus} }
   \end{minipage}
           \begin{minipage}[t]{0.079\linewidth}
       \vspace*{-2.3mm}
       \centerline{\scriptsize{ Defocus} }
   \end{minipage}
   \begin{minipage}[t]{0.079 \linewidth}
       \vspace*{-2.3mm}
       \centerline{\scriptsize{ Defocus }}
   \end{minipage}
   \begin{minipage}[t]{0.079 \linewidth}
       \vspace*{-2.3mm}
       \centerline{\scriptsize{  In focus} }
   \end{minipage}
           \begin{minipage}[t]{0.079\linewidth}
       \vspace*{-2.3mm}
       \centerline{\scriptsize{ Defocus} }
   \end{minipage} 
\begin{minipage}[t]{0.079 \linewidth}
       \vspace*{-2.3mm}
       \centerline{\scriptsize{ Defocus }}
   \end{minipage}
   \begin{minipage}[t]{0.079 \linewidth}
       \vspace*{-2.3mm}
       \centerline{\scriptsize{ In focus} }
   \end{minipage}
           \begin{minipage}[t]{0.079\linewidth}
       \vspace*{-2.3mm}
       \centerline{\scriptsize{ Defocus} }
   \end{minipage}
   % \begin{minipage}[t]{0.16 \linewidth}
   %     \vspace*{-2.3mm}
   %     \centerline{\footnotesize{  Mid focus } }
   %  \end{minipage}
   %  \begin{minipage}[t]{0.16 \linewidth}
   %     \vspace*{-2.3mm}
   %     \centerline{\footnotesize{ \hspace{0.2mm} Back focus} }
   %  \end{minipage}
    \end{minipage}

 \vspace*{-2mm}

      \caption{ 
Visual comparison on simulations between~\textcolor{myred}{ASM 6}  and~\textcolor{myblue}{Ours 6} at six depth planes under 0 mm and 10 mm propagation distances. (Source image:~Martin Kníže, Link:~\href{https://commons.wikimedia.org/wiki/File:Sleeping_tiger_up_close_(Unsplash).jpg}{Wikimedia Commons}) }
           
                 \label{fig:capture_simulation}
    \end{center}

\end{figure*}

% \vspace*{-6mm}
% \begin{figure}[h!]
%     \begin{center}
%     \vspace*{-2mm}
% \begin{minipage}[t]{1\linewidth} 
   
%     \begin{minipage}[t]{1\linewidth}
%        {\includegraphics[width=1\linewidth]{figures/experimentFigure/supp_capture.png}}
%     \end{minipage}

%        \begin{minipage}[t]{0.47\linewidth}
%        \vspace*{-2.3mm}
%        \centerline{\footnotesize{ \quad \quad \textcolor{myblue}{Ours 6}}}
%    \end{minipage}
%    \begin{minipage}[t]{0.47 \linewidth}
%        \vspace*{-2.3mm}
%        \centerline{\footnotesize{ \quad \quad \textcolor{myred}{ASM 6}  }  }
%     \end{minipage}
    
%     \end{minipage}
%          \vspace*{-4mm}    
%       \caption{ Comparing experimental captures of ~\textcolor{myred}{ASM 6} and~\textcolor{myblue}{Ours 6}  under 0 mm propagation distances.}
           
%                   \label{fig:capture}
%     \end{center}
% \end{figure}

\begin{wrapfigure}[7]{l}{14cm}
\centering
\vspace{-5mm}
\includegraphics[width=1\linewidth]{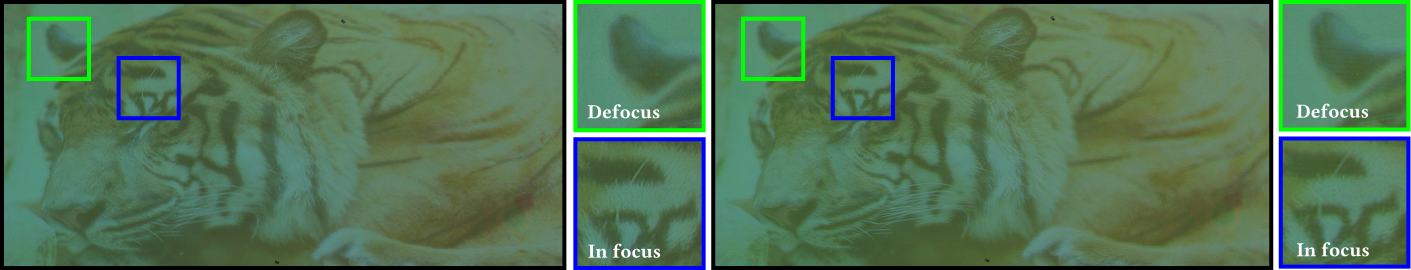}
       \begin{minipage}[t]{0.2\linewidth}
       \vspace*{-2.3mm}\hspace{-10mm}
       \centerline{\scriptsize{ \textcolor{myblue}{Ours 6}}}
   \end{minipage}
   \begin{minipage}[t]{0.45 \linewidth}
       \vspace*{-2.3mm}\hspace{12mm}
       \centerline{\scriptsize{ \quad \quad \textcolor{myred}{ASM 6}  }  }
    \end{minipage}
\hspace{-2mm}
\vspace{-2mm}
\caption{ Comparing experimental captures of ~\textcolor{myred}{ASM 6} and~\textcolor{myblue}{Ours 6}  under 0 mm propagation distance.~(Source image:~Martin Kníže, Link:~\href{https://commons.wikimedia.org/wiki/File:Sleeping_tiger_up_close_(Unsplash).jpg}{Wikimedia Commons})}
\label{fig:holographic_display}
\end{wrapfigure}